\documentclass[aps,prd,nofootinbib,showpacs,amssymb,twocolumn]{revtex4}
\usepackage{amsmath,amsfonts}
\usepackage{epsfig,subfigure}
\usepackage{ifpdf}
\usepackage{hyperref}
\usepackage{amsmath}
\usepackage{graphicx}
\usepackage{color}
\usepackage{natbib}
\usepackage{multirow}

\newcommand{\msun}{$M_\odot$}
\newcommand{\be}{\begin{equation}}
\newcommand{\ee}{\end{equation}}
\newcommand{\bea}{\begin{eqnarray}}
\newcommand{\eea}{\end{eqnarray}}

\newcommand\mc{ {{\cal M}_c}}

\begin{document}

\author{Hee-Suk Cho}
\affiliation{Department of Physics, Pusan National University, Busan 609-735, Korea}

\author{Chang-Hwan Lee}
\affiliation{Department of Physics, Pusan National University, Busan 609-735, Korea}

\title{Validity of the Effective Fisher matrix for parameter estimation analysis: Comparing to the analytic Fisher matrix}
\date{\today}

\begin{abstract}
The effective Fisher matrix method recently introduced by Cho {\it et al}.~\cite{Cho13} is a semi-analytic approach to the Fisher matrix, in which a local overlap surface is fitted by using a quadratic fitting function. Mathematically, the effective Fisher matrix should be consistent with the analytic one at the infinitesimal fitting scale.
In this work, using the frequency-domain waveform (TaylorF2), we give brief comparison results between the effective and analytic Fisher matrices for several non-spinning binaries consisting of binary neutron stars with masses of  (1.4, 1.4)\msun, black hole-neutron star of (1.4, 10)\msun, and binary black holes of (5, 5) and (10, 10)\msun  $ \ $for a fixed signal to noise ratio (SNR=20) and show a good consistency between two methods. We also give a comparison result for an aligned-spin black hole-neutron star binary with a black hole spin of $\chi=1$, where we define new mass parameters ($\mc, \eta^{-1}, \chi^{7/2}$) to find good fitting functions to the overlap surface. 
The effective Fisher matrix can also be computed by using the time-domain waveforms which are generally more accurate than frequency-domain waveform. We show comparison results between the frequency-domain and time-domain waveforms (TaylorT4) for both the non-spinning aligned-spin binaries.
\end{abstract}

\pacs{04.30.--w, 04.80.Nn, 95.55.Ym, 02.70.--c, 07.05.Kf}

\maketitle


\section{ Introduction} 
In gravitational wave data analysis, the parameter estimation methods are implemented to figure out the physical parameters of the wave source. 
The relevant information is the distribution of the measured values and the error bounds on their variances.
There are several methods for parameter estimation that are based on Monte Carlo simulations. 
They are able to search the whole parameter space but in general computationally very expensive.
One of those methods is the  Markov chain Monte Carlo (MCMC)~\cite{Cor06,Van09,Cor11,Vei12,Osh13}, which involves the Bayesian analysis framework.

The Fisher matrix method has been generally used to estimate the error bounds~\cite{Poi95,Aru05,Lan06,Bro07,Osh13}.
The inverse of the Fisher matrix represents the covariance matrix, from which the error bounds can be directly derived as well as the correlations between the parameters.
Although the Fisher matrix is valid only for the high signal to noise ratio (SNR), this method is very useful because that can compute measurement accuracies very quickly compared to the MCMC.

The Fisher matrix prediction has only used the frequency-domain waveform, so called TaylorF2 (or SPA), because of the possibility of deriving analytical expression of the Fisher matrix. 
The TaylorF2 waveform has only been applied for comparison between the MCMC and Fisher matrix so far~\cite{Cok08,Rod13}.
The frequency-domain waveforms require much less computational time for the MCMC runs.
However, the time-domain waveforms are basically more accurate because that do not assume the stationary phase approximation.
The MCMC methods have used various time-domain waveforms for more accurate parameter estimation performance~\cite{Van08,Van09,Cho13,Aas13,Osh13}.
Motivated by that, first and foremost, Cho {\it et al.}~\cite{Cho13} introduced an effective method, with which they calculated the Fisher matrices using the time-domain waveform, TaylorT4.
When using the time-domain waveforms for the Fisher matrix, it may be very complicated to obtain the derivatives of the waveforms because the Fourier transform of the time-domain waveform is not the analytic function but numerical data.
The effective method can avoid the difficulty by fitting the local overlap surface, where the Fisher matrix can be derived from the quadratic fitting function.

The analytic Fisher matrix method is straightforward because the result can be derived analytically from the analytic waveform function except for overlap integration.
While, the effective method involves the fitting function which is manually calculated, and the overlap integration can be done by the same manner as in the analytic method.
These two methods should give the same results under certain physical conditions.
Therefore, in order to accept the effective Fisher matrix generally, the faithfulness of that should be proved by comparing to the analytic Fisher matrix.
Several works~\cite{Cor06,Cok08,Rod13} showed inconsistency between the Fisher matrix and MCMC results. However, the authors emphasize that
the main purpose of this work is not to prove the adequacy of the Fisher matrix in parameter estimation analysis but to investigate the possibility of the effective method to calculate the Fisher matrix.

In Sec.~II, we review the TaylorF2 waveform and the overlap formalism. We review the analytic and effective Fisher matrix methods in Sec.~III.
We show our comparison results between the analytic and effective methods as well as the results between the TaylorF2 and TaylorT4 waveforms.
In Sec.~V, we summarize our results and give some discussions. 
Throughout this paper, all mass parameters are in units of the solar mass (\msun) unless otherwise noticed,  and we use a geometrized unit, where $G=c=1$ .


\section{Wave function and overlap} 
The TaylorF2 waveform is given by
\be
\tilde{h}(f)=Af^{-7/6}e^{i\Psi(f)},
\ee
where $A \propto \mc^{5/6}\Theta(\rm{angle})/D$, $\mc$ is a chirp mass, $D$ is the luminosity distance of the binary, and $\Psi(f)$ is the orbital phase. $\Theta(\rm {angle})$ is a function of the orbital orientation with respect to the detector network in terms of the sky position (RA, DEC), orbital inclination ($\iota$), and the wave polarization ($\psi$).
If we assume the fixed SNR of the waveforms, all information of the waveform is coming from the wave phase. The phasing factor consists of the coalescence time ($t_c$) and termination phase ($\phi_0$), and the remaining intrinsic parameters ($\lambda_{\rm int}$):
\be\label{eq.phasing}
\Psi(f)=2\pi f t_c - 2 \phi_0 - {\pi \over 4} + {3 \over 128 \eta}F(\lambda_{\rm int}, f),
\ee
where  $t_c$ can be chosen arbitrarily, $F(\lambda_{\rm int}, f)$ can be represented by the post-Newtonian (pN) expansion, which is provided in \cite{Aru05} up to 3.5 pN order for the non-spinning case, where $\lambda_{\rm int}=\{\mc, \eta\}$, $\eta$ is a symmetric mass ratio.
For the aligned-spin case, a dimensionless spin parameter $\chi$ is included, so  $\lambda_{\rm int}=\{\mc, \eta, \chi\}$, and $F(\lambda_{\rm int}, f)$  is provided in \cite{Cut94, Poi95, Aru09}.

The termination phase ($\phi_0$) is related to the coalescence phase ($\phi_c$) by~\cite{Sat91, All12}
\be \label{eq.phi0}
2\phi_0=2\phi_c-{\rm arctan}\bigg({F_{\times} \over F_+ }{2 \cos \iota \over 1+\cos^2\iota}\bigg),
\ee
where $F_{\times}$ and $F_+$ are the antenna response functions depending on the angle parameters ($\iota, \psi$, RA, DEC).
For simplicity, we consider a fixed binary position, then $\phi_0$ is a function of $\iota, \psi$, and the coalescence phase $\phi_c$ (the coalescence phase can also be chosen arbitrarily). 
In several works (e.g., \cite{Cut94,Poi95}) $\phi_0$ has been assumed to be an arbitrary constant when calculating Fisher matrices, and they considered only one angle parameter as components in the Fisher matrices. However, in order to take into account more than two angle parameters among ($\iota, \psi, \phi_c$), one should define the $\phi_0$ as a function of the angle parameters ($\iota, \psi, \phi_c$). For example, if the binary is optimally placed and orientated (i.e., $\iota=$RA$=$DEC$=0$), the phase $\phi_0$ is exactly degenerated into $\phi_c$ and $\psi$ by a function of $\phi_0=\phi_c - \psi$. In this case, the Fisher matrix is singular and the inverse matrix can not be defined. The correlation between these two parameters becomes reduced as the $\iota$ increases.
If $\iota=\pi/2$, $\phi_0$ is equal to the arbitrary constant $\phi_c$ and other angle parameters can be removed from the wave phase equation.
In this work, we only consider the fixed binary orientation, so $\phi_0$ is assumed to be the same as in the previous works, then the wave phase in Eq.~(\ref{eq.phasing}) is determined by a combination of the parameters ($\lambda_{\rm int}, t_c, \phi_c$).
However, when computing the analytic Fisher matrix which considers both $\phi_c$ and $\psi$ with other physical parameters of interest, one should not set $\phi_0$ equal to $\phi_c$ in general.

The overlap between a signal ($h_s$) and a template ($h_t$) is defined by 
\be \label{eq.conventionaloverlap}
\langle h_s | h_t \rangle = 4 {\rm Re} \int_{0}^{\infty}  \frac{\tilde{h_s}(f)\tilde{h}_t(f)^*}{S_n(f)} df,
\ee
where $\tilde{h}(f)$ is the Fourier transform of $h(t)$. Note that the inverse Fourier transform will compute the overlap for all possible coalescence times of $h_t$ at once~\cite{All12}. In addition, by taking the absolute value of the complex number we can maximize the overlap over all possible coalescence phases~\cite{All12}, 
\be\label{eq.overlap}
{\rm max}_{t_c,\phi_c}\langle h_s | h_t \rangle \equiv 4 \bigg| \int_{0}^{\infty}  \frac{\tilde{h_s}(f)\tilde{h}_t(f)^*}{S_n(f)}  e^{2 \pi i f t} df\bigg |.
\ee
Finally, the normalized overlap is defined by
\be\label{eq.maxoverlap}
P(h_s,h_t) =  {\rm max}_{t_{\rm c},\phi_c}{\langle h_s | h_t \rangle   \over  \sqrt{\langle h_s | h_s \rangle \langle h_t | h_t \rangle }} .
\ee
In Eq.~(\ref{eq.overlap}), we assume the analytic initial LIGO sensitivity curve~\cite{Dam01,Aji09}, which takes the form:
\bea\label{eq.noise}
S_h(f)&=&S_0\bigg[\bigg( {4.49 f \over f_0}\bigg)^{56} \\ 
& +& 0.16 \bigg( {f \over f_0}\bigg)^{-4.52} + 0.52 + 0.32\bigg( {f \over f_0}\bigg)^2   \bigg) \bigg], \nonumber
\eea
where $f_0=150$ Hz, and $S_0=9 \times 10^{-46}$ ${\rm Hz}^{-1}$.


\section{Effective Fisher matrix: semi-analytic approach} 
The Fisher matrix is defined by
\be \label{eq.analyticfisher}
\Gamma_{ij}=\bigg\langle {\partial h \over \partial \lambda_i} \bigg | {\partial h \over \partial \lambda_j} \bigg \rangle,
\ee
where $\lambda_i=\{\lambda_{\rm int}, \phi_c, t_c\}$. In this equation, since the TaylorF2 is an analytic function, derivatives with respect to the parameters can be analytically computed, and the Fisher matrix can be calculated by integrating derivatives of two functions numerically. The analytic results of Fisher matrices in this work are
computed by using the software package {\it Mathematica}. 

On the other hand, a Fourier transform of the  time-domain waveform is only numerical data. In this case, it is very complicated to obtain the derivatives, so almost impossible to compute the Fisher matrix.  
However, in Eq.~(\ref{eq.analyticfisher}), the derivatives involves with respect to the physical parameters, while the overlap integral with respect to the frequency, therefore, both computations are formally commutable.
Then, the Fisher matrix can be directly derived by the log likelihood ($\ln L$), and the log likelihood is a function of the overlap ($P$) \cite{Jar94,Val08,Cho13}:
\be
\Gamma_{ij}=-{\partial^2 \ln L(\lambda) \over \partial \lambda_{i} \partial \lambda_{j}}
                          =\rho^2  {\partial^2 (1-P) \over \partial \lambda_{i} \partial \lambda_{j}}\bigg|_{\lambda_i,\lambda_j=\lambda_0},
\ee
where $\lambda_0$ is the fiducial value of each parameter and $\rho$ is the SNR. 
Using this expression, one can compute the Fisher matrix by differentiating the overlap with respect to the corresponding parameters at the position of the fiducial values of parameters.
We still have to calculate the derivatives of the overlap, which is also numerical data. 
However, for Gaussian noise and high SNR, since the likelihood function is a normal distribution assuming a flat prior~\cite{Fin92,Cut07}, a quadratic fitting function best fits the log likelihood. So if we find an analytic fitting function ($P^*$) at a physically appropriate scale, the derivatives of the fitting function can be analytically obtained.
Using this semi-analytic method, Cho {\it et al.}~\cite{Cho13} introduced the effective Fisher matrix as
\be
\Gamma_{\rm eff}=-{\partial^2 P^* \over \partial \lambda_{i} \partial \lambda_{j}}\bigg|_{\lambda_i,\lambda_j=\lambda_0} \simeq \Gamma_{ij}/\rho^2.
\ee
Note that in the infinitesimal fitting scale, the effective Fisher matrix exactly reflects the analytic result in Eq. (\ref{eq.analyticfisher}). In this work, we choose very localized overlap surfaces in the range of $P>0.999$.

It has been found that the inverse of the Fisher matrix is the covariance matrix ($\Sigma_{ij}$), and the measurement error ($\sigma_i$) of each parameter and correlation coefficient ($c_{ij}$) between two parameters are obtained by
\be
\sigma_i=\sqrt{\Sigma_{ii}}, \ \ \ c_{ij}={\Sigma_{ij} \over \sqrt{\Sigma_{ii}\Sigma_{jj}}}.
\ee
Note that the covariance matrix is a inverse of the Fisher matrix (not the effective Fisher matrix), so depends on both the SNR and effective Fisher matrix as $\Sigma_{ij}=(\rho^2 \Gamma_{\rm eff})^{-1}$, consequently the parameter estimation error also depends on SNR simply as $\sigma_i(\rho)=\sigma_i(1)/\rho$. In this work, we assume the SNR to be $\rho=20$. The correlation coefficient ($c_{ij}$) is $\rho$-independent.


\section{Result: comparison to the analytic Fisher matrix} 
In this work, we take into account four non-spinning binary models, whose mass components are summarized in Table.~\ref{tab.parameters}.
\begin{table}[t]
\begin{tabular}{ccccc}
                                     &BNS& BHNS&BBH1&BBH2 \\

\hline
$m_1, m_2$  & 1.4, 1.4&10, 1.4&5, 5&10, 10 \\
\hline
$\mc$&1.2188&2.9943&4.3528&8.7055\\
$\eta$&0.25&0.1077&0.25&0.25\\
\hline
$f_{\rm LSO}$ [Hz]&1570&386&440&220\\
\hline
$T_{\rm chirp}$ [s]&53.54&11.97&6.42&2.02\\
\end{tabular}
\caption{\label{tab.parameters}{\bf Fiducial values of the mass parameters for the non-spinning binaries.} The frequencies at the last stable orbit ($f_{\rm LSO}$) are calculated by Eq.~(\ref{eq.flso}), the ``chirp time" $T_{\rm chirp}$ is determined by Eq.~(\ref{eq.chirptime}).}
\end{table}
For the aligned-spin case, we consider the BHNS binary with a black hole spin of $\chi=1$ where $\chi$ is a dimensionless spin parameter, so the black hole is maximally rotating.

We use the TaylorF2 waveform that is implemented in the LIGO Algorithm Library (LAL) \cite{lal} to calculate the overlap surfaces for the effective Fisher matrices. We consider only the restricted waveforms with the phase term up to 3.5 pN for both the non-spinning and aligned-spin waveforms, where only the 1.5 pN spin-orbit phase correction is contained in the aligned-spin case.
The maximum frequency cutoff for the TaylorF2 waveform is taken when the binary hits the ``last stable orbit (LSO)". For simplicity, we assume the expression for a test particle orbiting a Schwarzschild black hole with a mass of $M=m_1+m_2$ for both the non-spinning and aligned-spin cases:
\be \label{eq.flso}
f_{\rm LSO}={1 \over 6^{3/2}\pi M}.
\ee

In many previous works, the minimum frequency has been taken to be 40 Hz according to the noise sensitivity curve, 
but we choose 30 Hz not to lose any additional information in the lower frequency region. O'Shaughnessy {\it et al.}~\cite{Osh13} showed a non trivial difference between the minimum frequencies of 30 and 40 Hz (e.g., see figure. 8 theirin) using the time-domain waveform, TaylorT4. We also give some similar results for the TaylorF2 waveform including a dependence on the upper frequencies in Appendix A.

The multivariate covariance matrix is very sensitive to the components of Fisher matrix, so high precision numbers in the Fisher matrix should be presented to produce the exact inverse Fisher matrix.
In this work, it is more convenient to represent the covariance matrices instead of Fisher matrices. Exact components of the Fisher matrix can be determined by an inverse of the covariance matrix.

\subsection{Non-spinning binaries} 
\begin{figure}[!]
\includegraphics[width=\columnwidth]{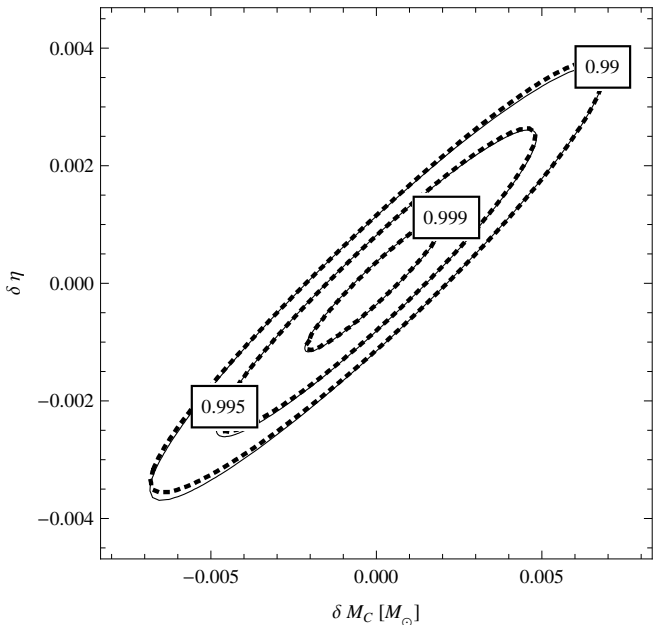}
\caption{{\bf Comparison between overlap contours and corresponding quadratic fittings for BHNS binary, showing the quadratic overlap surface.}  ${\bf \delta \lambda}$ is defined by a difference from the fiducial value in the Table~\ref{tab.parameters}. The dotted lines correspond to the overlap contours and the solid lines correspond to the fitting ellipses using eq.(\ref{eq.fitting}).\label{fig.fitting}}
\end{figure}

As stated in the previous section, we find that the local overlap surfaces are almost quadratic  at the region of $P > 0.99$,  so  the Fisher matrix does not depend on the fitting scale. As an example, for the BHNS binary the fitting function is
\be \label{eq.fitting}
P^*=1-[1738 (\delta \mc)^2 - 64978 (\delta \mc) (\delta \eta) + 6676 (\delta \eta) ^2].
\ee
Figure.~\ref{fig.fitting}  shows the overlap contours  and corresponding fitting ellipses using this fitting function. At the scale of $P > 0.999$ the overlap surface is almost exactly quadratic.
In this work, unless otherwise noted, we assume the fitting scale of $P>0.999$ to calculate the fitting functions ($P^*$).

In Table.~\ref{tab.noSpin}, we summarize the comparison results between the effective and analytic Fisher matrices for the non-spinning binary models.
For all binary models, we find a very good agreement between two methods. Especially, for the BNS and BHNS models the correlations and measurement errors for both methods are
almost exactly the same.

On the other hand, although we assumed the same SNR ($\rho=20$) for all binary systems, the fractional errors of parameter measurement are distributed quite broadly by ranges of $0.01-0.7 \%$ and $0.8-5.5 \%$ for $\mc$ and $\eta$, respectively. We see that the accuracy is related with the ``chirp time" of the binary system, which is the amount of time that the binary system will take from a minimum frequency ($f_{\rm min}$) to coalescence. At the leading pN order, the chirp time is given by  \cite{All12} 
\be \label{eq.chirptime}
T_{\rm chirp} = \frac{5}{256}\frac{M(\pi f_{\rm min}M)^{-8/3}}{\eta}= \frac{5}{256}\frac{(\pi f_{\rm min})^{-8/3}}{ \mc^{5/3}} .
\ee
The total number of cycles of a binary is proportional to the chirp time for a given $f_{\rm min}$ \cite{Aru05}:
\bea
N_{\rm cycle}&=&{1 \over 32}\frac{(\pi f_{\rm min} M)^{-5/3}}{\pi \eta}={1 \over 32}\frac{f_{\rm min}(\pi f_{\rm min})^{-8/3}}{\mc^{5/3}}  \nonumber \\
&=& {8 \over 5}f_{\rm min}T_{\rm chirp} .
\eea
Roughly speaking, as the the time of inspiral is longer, the more number of cycles the signal has, and the more information of a wave signal can be accumulated,
consequently the measurement accuracy can be improved.
Note that, the chirp mass is a function of not only a total mass but also a symmetric mass ratio, so the higher $f_{\rm max}$ does not always mean the longer $T_{\rm chips}$. For example, the total mass of the BHNS is similar to that of BBH1 but the chirp time is about twice longer, giving smaller errors by a factor of a half for both $\mc$ and $\eta$.

To investigate the correlation of parameter estimation performance with detector characteristics, Arun {\it et al.}~\cite{Aru05} considered the number of ``useful cycles" instead of the $N_{\rm cycle}$. They showed various results using advanced LIGO, initial LIGO (where $f_{\rm min}=40$ Hz), and VIRGO detectors for NSNS, BHNS, and BBH2 models varying post-Newtonian orders of the TaylorF2 waveform. In addition to these three models, Cokelaer \cite{Cok08} also provided error for a BBH2 system assuming the initial LIGO sensitivity curve (where $f_{\rm min}=40$ Hz).

\subsection{Aligned-spin BHNS binary} 
We select the BHNS binary model in Table.~\ref{tab.parameters} for the case of aligned-spin binary.
In general, the NS spin can be neglected, and we need one additional parameter $\chi$ in the function of waveforms, 
so the number of intrinsic parameters of interest is three, then the fitting function should be a 3-dimensional ellipsoid.
We assume a maximally rotating black hole ($\chi=1$), then a shape of the overlap surface is a very long and thin ellipsoid.
We find, unfortunately, the surface is not symmetric even at the fine scale of $P=0.999$.
Since we want to fit the overlap surface quadratically, we go into the very localized region where the overlap surface becomes symmetric.
We find, that is enough symmetric at the scale of $P>0.99999$. For exact fitting at this scale, however, 
very high sampling rates are necessary in the overlap integration to reduce the numerical errors (for more explanations, refer to Sec. III-G of \cite{Cho13}), which comes from maximizing over $\phi_c$ and $t_c$.
That means we need longer computing time for smaller fitting scales.

\begin{figure}[t]   
\includegraphics[width=4.2cm, height=4.2cm]{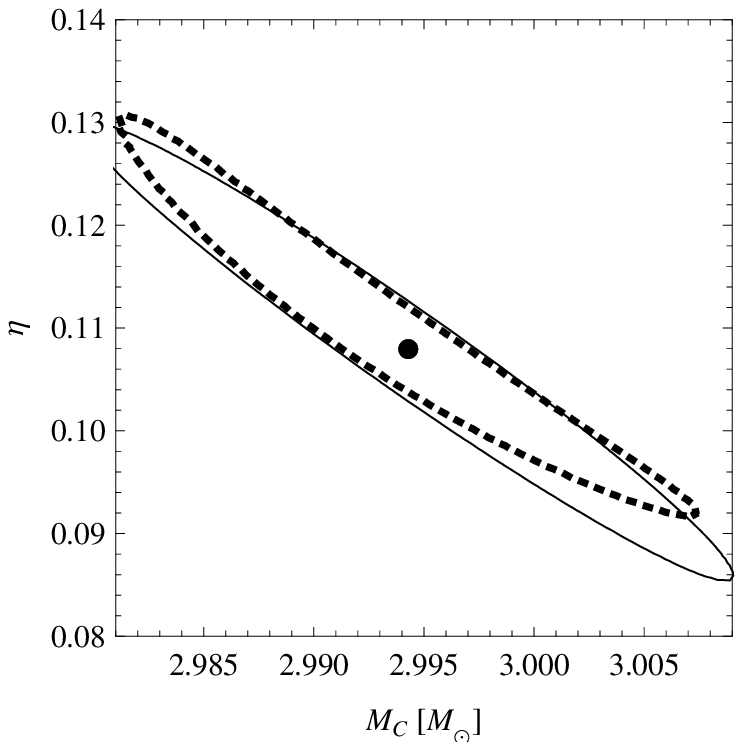}
\includegraphics[width=4.2cm, height=4.2cm]{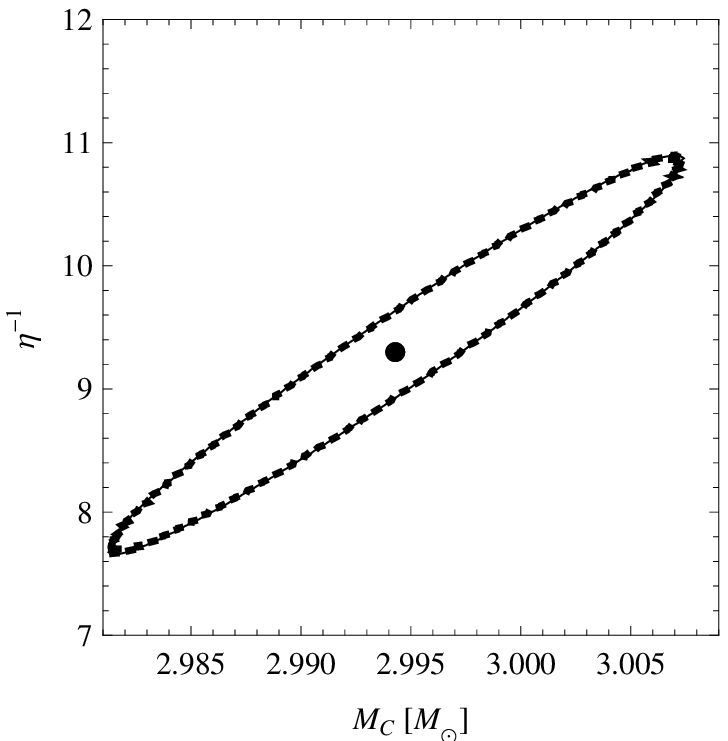}
\includegraphics[width=4.2cm, height=4.2cm]{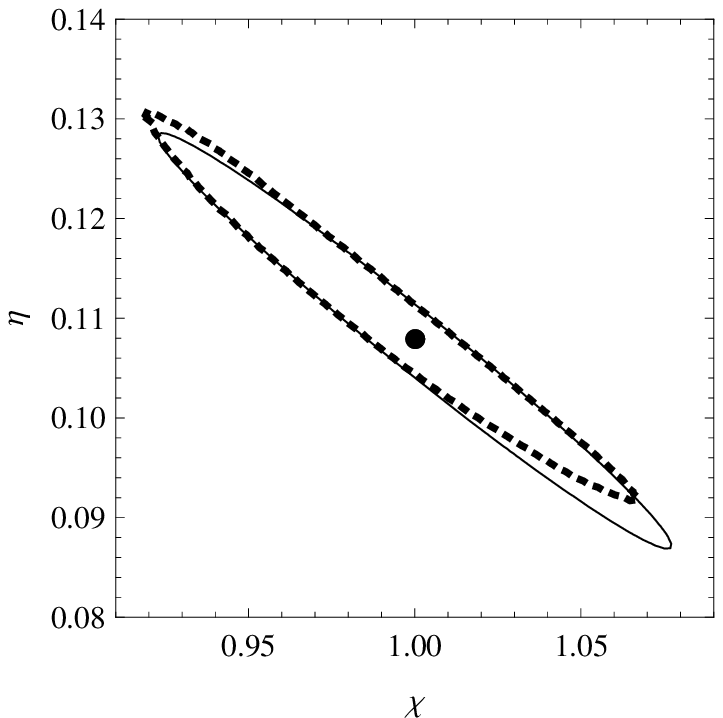}
\includegraphics[width=4.2cm, height=4.2cm]{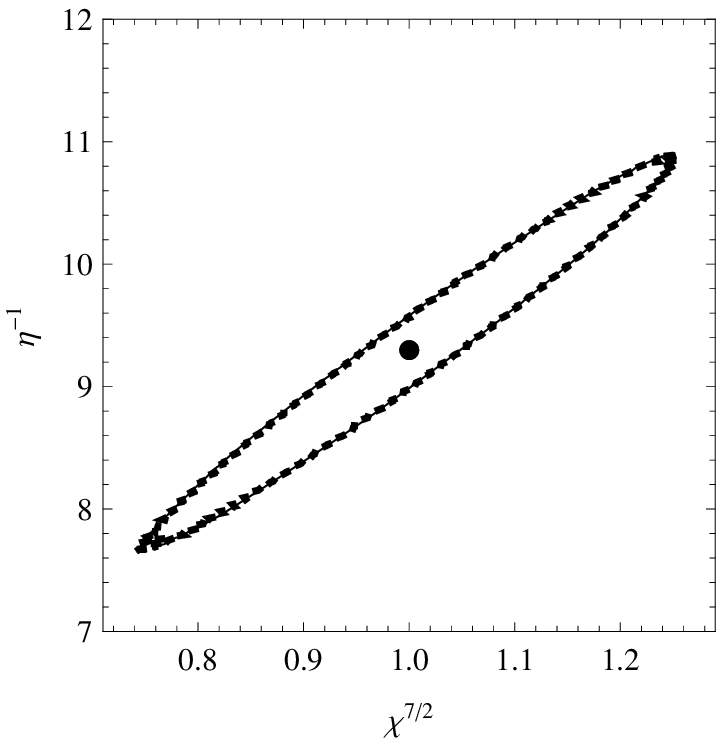}
\includegraphics[width=4.2cm, height=4.2cm]{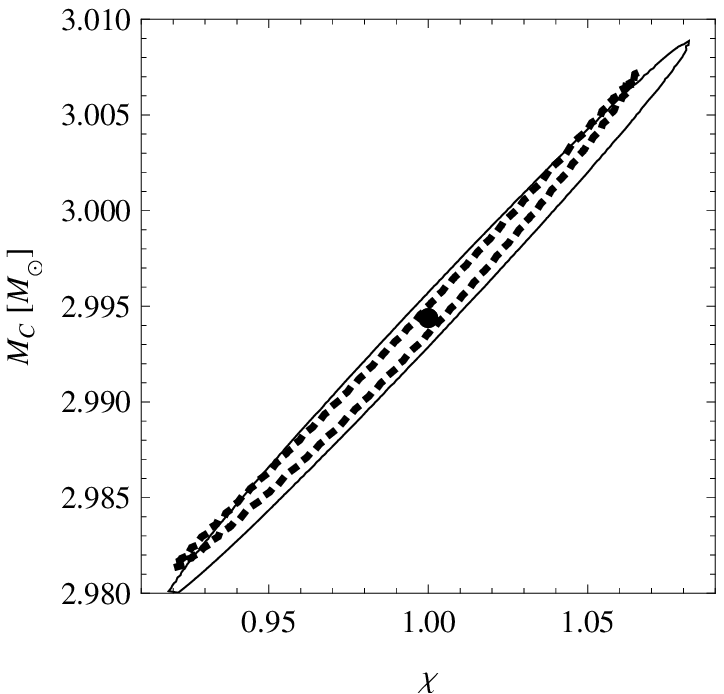}
\includegraphics[width=4.2cm, height=4.2cm]{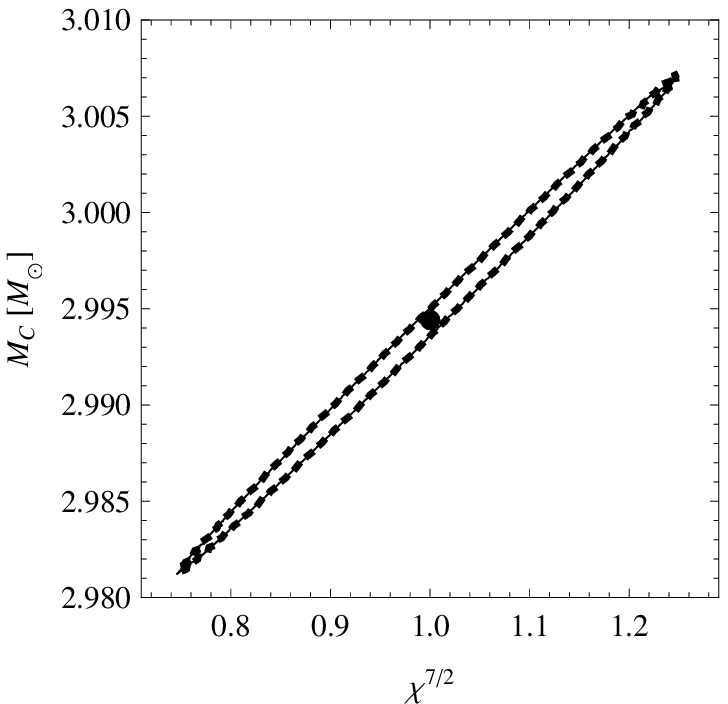}
\caption{\label{fig.alignedSpin}{\bf Projections of the overlap ellipsoid ($P=0.999$) and their fitting functions for the aligned-spin BHNS binary.} Dotted lines indicate the overlap contours and solid lines correspond to their quadratic fitting ellipses. Large dots indicate the fiducial parameter values. Note that the fitting functions are in good agreement with the overlap contours when using the new parameters ($\mc, \eta^{-1}, \chi^{7/2}$).}
\end{figure}

On the other hand, fortunately, by choosing a new parameter coordinate, the overlap surface can be symmetric at the scale of $P>0.999$.
We found the effective parameter coordinate by converting as $\eta \rightarrow \eta^{-1}$ and $\chi \rightarrow \chi^{7/2}$.
Fig.~\ref{fig.alignedSpin} shows projections of the overlap ellipsoid, where $P=0.999$, represented in the ($\mc, \eta, \chi$) and ($\mc, \eta^{-1}, \chi^{7/2}$) coordinates.
Like the case of non-spinning binary in Fig.~\ref{fig.fitting}, the fitting function is in good agreement with the overlap surface by using the new coordinate.
In Table.~\ref{tab.alignedSpin}, we give the results of the effective method at different scales, and using different coordinates of ($\mc, \eta, \chi$) and ($\mc, \eta^{-1}, \chi^{7/2}$), also give the comparison
to the analytic results.
We can get a more exact fitting function by decreasing the fitting scale. But, by using new parameters instead of decreasing the fitting scale, the consistency between the effective and analytic methods can be improved. 
Using the new parameters, the results for both methods are in perfect agreement at the scale of $P>0.999$.

Note that, since we found this new parameters empirically, that may not work for other binary models.
Nevertheless, choosing the new coordinate is good approach for the purpose of this work, which is to find quadratic fitting functions to the overlap surface.

\subsection{Comparison to the time-domain waveform}
O'Shoughnessy {\it et al.}~\cite{Osh13} investigated the effective Fisher matrices using  the time-domain waveform, TaylorT4 implemented in the LAL~\cite{lal}, for the non-spinning and aligned-spin BHNS binaries.
They used the initial LIGO sensitivity curve and assumed $f_{\rm min}=30$ Hz as the same conditions as in this work.
We summarize some of their results with our results for comparison of Fisher matrices between the TaylorF2 and TaylorT4 waveforms in Table.~\ref{tab.fversust}.
We find small differences between TaylorF2 and TaylorT4 waveforms for both the non-spinning and aligned-spin binaries.

The TaylorT4 waveform is terminated when the binary reaches the ``minimum energy circular orbit (MECO)" \cite{ Bla02,Buo03}, so $f_{\rm max}=f_{\rm MECO}$. With this waveforms, the MECO frequencies are different depending on the BH spin. For example, $f_{\rm MECO} \sim 560$ and $\sim 890$ Hz for the non-spinning and aligned-spin (where only the 1.5 pN spin-orbit phase correction is contained) BHNS binaries, respectively.
For exact comparison, in Table.~\ref{tab.fversust}, we set the $f_{\rm max}$ of the TaylorF2 waveform to be the same frequency as the $f_{\rm MECO}$ of TaylorT4 waveform instead of $f_{\rm LSO}$.
Note that, since we have already shown good consistency between the effective and analytic Fisher matrices for the BHNS binaries, we used the analytic method for the TaylorF2 waveform in this Table.


\section{Summary and discussion} 
In this work, we have investigated the effective and analytic Fisher matrices using TaylorF2 waveform for various non-spinning binary models and an aligned-spin BHNS binary. 
The effective Fisher matrix is computed by differentiating the fitting function to the local overlap surface, so mathematically that should be the same as the analytic Fisher matrix at the infinitesimal fitting scale.
We have shown that the two methods are in very good agreement at the scale of $P>0.999$ for all binary models concerned in this work, introducing a new coordinate ($\mc, \eta^{-1}, \chi^{7/2}$) for the aligned-spin case.
We have also shown some results for the time-domain waveform, TaylorT4, and given comparison to the TaylorF2 results for the non-spinning and aligned-spin BHNS binaries.
We have found small differences between TaylorF2 and TaylorT4 waveforms.

For the time-domain waveforms, the effective method can avoid the inconvenience in computing the derivatives of the numerical waveform data.
The effective method simply computes the derivatives by fitting the local overlap surface.
The best advantage of the effective method is that we can easily compute the Fisher matrix using various time-domain waveforms implemented in LAL.
For more accurate performances of the parameter estimation analysis, the MCMC methods have used the time-domain waveforms, that require very long computational time depending on the binary model.
By using the effective Fisher matrices, however, we can investigate the parameter estimate performances for various binary models prior to the real MCMC runs.
Waveform models that contain the merger-ring down phase, can also be usable for the effective Fisher matrix.

The measurement errors derived from the analytic Fisher matrix is independent of the SNR, that just fall off as the inverse of SNR.
This behavior is similar for the effective Fisher matrix if we use the TaylorF2 waveform.
When using the time-domain waveforms, however, one can investigate the dependence on SNR by choosing the fitting scales physically adjusted.
Cho {\it et al.}~\cite{Cho13} found that the fitting functions are pretty dependent on the SNR (e.g., see Fig. 2 theirin).
Their subsequent work~\cite{Osh13} showed good consistency of the effective Fisher matrices with MCMC parameter estimation results for the non-spinning and aligned-spin BHNS binaries using TaylorT4 waveforms with the SNR of 20.

For the precessing binary cases, time-domain waveforms have been developed and already implemented in the LAL \cite{lal}.
Cho {\it et al}.~\cite{Cho13} showed preliminary results of the effective Fisher matrices for precessing BHNS binaries. 
For comparison to their results, the MCMC runs are on going.
Recently, a frequency-domain waveform for the precessing cases was developed by \cite{Lun13} and implemented in the LAL,
that may be available for an extension of this work.

\begin{table*}[!]
\begin{tabular}{c | ccc|cc|cc|cc|cc|cc|cc|cc  }
  Binary                                                                          & \multicolumn{5}{|c}{BNS}       &  \multicolumn{4}{|c}{BHNS}  & \multicolumn{4}{|c}{BHBH1}  & \multicolumn{4}{|c}{BHBH2}    \\
     \hline
 Method       &&\multicolumn{2}{c|}{Effective}  &\multicolumn{2}{|c|}{Analytic}  &\multicolumn{2}{|c|}{Effective}  &\multicolumn{2}{|c|}{Analytic} 
 &\multicolumn{2}{|c|}{Effective}  &\multicolumn{2}{|c|}{Analytic} &\multicolumn{2}{|c|}{Effective}  &\multicolumn{2}{|c}{Analytic} \\
 \hline
 Parameter&&$\mc$ & $\eta$  &   $\mc$   &  $\eta$      &$\mc$ &  $\eta$    &$\mc$ &  $\eta$
                                                                             &$\mc$ & $\eta$  &   $\mc$   &  $\eta$      &$\mc$ &  $\eta$    &$\mc$ &  $\eta$\\ 
                                                                             \hline
$\Sigma_{ij} \times 10^6$& $\mc$ &  0.02638 &0.03212  &0.02638 &0.03219  & 5.704 &2.929     &5.692 &2.925 &41.47 &3458 &41.87 &3379 &3379 &742.7 &3615 &800.9    \\
                                              & $\eta$& -        &0.04501  &-        &0.04519   &       -&1.667     &-       &1.666 &-                         &31.75&-       &32.65&-       &174.2&-    &188.6      \\
 \hline
            $c_{ij}$                                         & $\mc$&  1.00&0.932&1.00  &0.932   &1.00&0.950&1.00 &0.950&1.00&0.953&1.00  &0.954&1.00   &0.970&1.00  &0.970   \\
                                                                   &$\eta$&-&1.00&-&1.00&-&1.00&-&1.00&-&1.00&-&1.00&-&1.00&-&1.00\\
                                                                                                                                      \hline
$\sigma_i \times 10^3$		           & &0.162  &2.12      &0.162  &2.13&       2.39  &1.29&2.39  &1.29&6.44  &5.63&6.47  &5.71&58.1  &13.2&60.1  &13.7\\
 \hline
$\Delta \sigma_i / \lambda_i \ [\%]$		           & &0.0133  &0.849      &0.0133  &0.850&       0.0798  &1.20&0.0797  &1.20&0.148  &2.25&0.149  &2.29&0.668  &5.28&0.691  &5.49\\

    \end{tabular}
 \caption{\label{tab.noSpin}{\bf Comparison between the effective and analytic Fisher matrices for the non-spinning binaries.} We assume $\rho=20$, then $\Sigma=(\rho^2 \Gamma_{\rm eff})^{-1}, \sigma(\rho)=\sigma(1)/\rho$ and $c_{ij}$ is $\rho$-independent. The frequency range for the overlap integral is [$30, f_{\rm LSO}$] Hz, where each number of $f_{\rm LSO}$ is presented in Table.~\ref{tab.parameters}. For all binary models, two methods show a very good agreement. Due to the differences of ``chirp time", the errors are distributed quite broadly between the binary models. For comparison, we give fractional errors ($\Delta \sigma_i / \lambda_i$).}
 \end{table*}

\begin{table*}[!]
{\footnotesize
\begin{tabular}{c | cccc|ccc|ccc||cccc|ccc  }
Method            &\multicolumn{4}{|c|}{Effective}  &\multicolumn{3}{|c|}{Effective ($P>0.99999$)}&\multicolumn{3}{|c||}{Analytic}   &\multicolumn{4}{|c|}{Effective}  &\multicolumn{3}{|c}{Analytic}  \\
     \hline
Parameter          &                         &$\mc$ &  $\eta$  &  $\chi$     &    $\mc$   & $\eta$  &  $\chi$ &    $\mc$   & $\eta$  &  $\chi$       &    & $\mc$   & $\eta^{-1}$  &  $\chi^{7/2}$          &$\mc$   &  $\eta^{-1}$  &  $\chi^{7/2}$   \\ 
 \hline
$\Sigma_{ij} \times 10^6   $ &  $\mc$ &273.2 & -433.2 & 1605  &223.8 & -316.9 & 1244 & 221.7 & -310.7 & 1226&  $\mc$ &      221.8 & 2679$\times 10$ & 4292         &221.7 & 2677$\times 10$ & 4291 \\
                                                        & $\eta$   & -        &712.8 & -2565 & -        &470.7 & -1777  &    -&457.0 & -1733 &        $\eta^{-1}$     &-    &3395$\times 10^3$ & 5228$\times 10^2$           &                 -&3393$\times 10^3$ & 5227$\times 10^2$ \\
			                            &  $\chi$     &   -        & -                 & 9461  &   -        & -                 &6936  & -          & -   &6801      &  $\chi^{7/2}$ &-            &-         & 8331$\times 10$       & -           &      -          &8332$\times 10$    \\
 \hline
$c_{ij}$            &$\mc$&1.00  &-0.982 &0.998&1.00  &-0.976 &0.998    &1.00&-0.976&0.998        & $\mc$ &     1.00  &0.976&0.998&               1.00&0.976&0.998 \\
                        & $\eta$&-          &1.00    &-0.988 &-          &1.00    &-0.983 & -      & 1.00&-0.983         & $\eta^{-1}$ &-                  &1.00&0.983&                   -     & 1.00& 0.983  \\
		&  $\chi$&-         &-             &1.00     &-         &-             &1.00    & -     &-   & 1.00        &  $\chi^{7/2}$   &        -       &-       &1.00&                        -   & -        &1.00   \\
\hline
$\sigma_i \times 10^3$ &       & 16.5  &26.7 &97.3&15.0  &21.7 &83.3  &14.9  &21.4 & 82.5  &   &        14.9 &1840&289&           14.9  &1840 & 289     \\		                                                  
    \end{tabular}
    }
 \caption{\label{tab.alignedSpin}{\bf Comparison between the effective and analytic Fisher matrices for the aligned-spin BHNS binary}. We assume $\rho=20$.  The maximum frequency is taken to be $f_{\rm LSO}$ (386 Hz). For a  smaller fitting scale, we need higher sampling rates for the overlap integration, consequently longer computing time. Instead, by choosing a new coordinate ($\mc, \eta^{-1}, \chi^{7/2}$), consistency between the effective and analytic methods can be improved  at the scale of $P>0.999$.}
 \end{table*}

\begin{table*}[!]
\begin{tabular}{c | ccc|cc|ccc|ccc  }
   Source                                                                          & \multicolumn{5}{|c}{Non-spinning}                                                                    & \multicolumn{6}{|c}{Aligned-spin}   \\
     \hline
Waveform                                                               &\multicolumn{3}{|c|}{TaylorF2}  &\multicolumn{2}{|c|}{TaylorT4}   &\multicolumn{3}{|c|}{TaylorF2}  &\multicolumn{3}{|c}{TaylorT4}  \\
     \hline
Parameter                                                                   &                         &$M_{\rm c}$ &  $\eta$  &   $M_{\rm c}$   &  $\eta$      &$M_{\rm c}$ &  $\eta$ & $\chi$          &$M_{\rm c}$ &  $\eta$ & $\chi$  \\ 
 \hline
$\Sigma_{ij}   \times 10^6$                                                                &  $M_{\rm c}$ &4.186 &1.915  &5.252 &3.076& 75.48 & -79.82 & 406.4&68.44 & -49.42 & 344.9   \\
                                                                                                                       & $\eta$   & -         &0.994    &-        &1.966   &       -    &91.81 & -439.3  &           -     &40.47 & -256.4\\
			                                                                                            & $\chi$&   -        & -            & -          & -           &          -    & -       & 2208      & -          &      -          &1756    \\
 \hline
$c_{ij}$                                                                    &$M_{\rm c}$&1.00  &0.939    &1.00& 0.957&          1.00  &-0.959&0.996&              1.00  &-0.939 &0.995 \\
                                                                                             & $\eta$&-          &1.00     & -      & 1.00&         -         &1.00&-0.976   &                   -     &1.00     &-0.962 \\
			                                                                   & $\chi$&-         &-             &      -     & -        &               -       &-       &1.00&                        -   & -        &1.00  \\
\hline
$\sigma_i \times 10^3$			                                       &       & 2.05  &0.997 &             2.29 &1.40&         8.69  &9.58 & 47.0    & 8.27   &6.36 &41.9 \\	
    \end{tabular}
 \caption{\label{tab.fversust}{\bf Comparison between the TaylorF2 and TaylorT4 waveforms for the BHNS binaries}: We assume $\rho=20$. We only consider  the 1.5 pN spin-orbit phase correction for the aligned-spin case. The results for TaylorF2 are analytically computed and those for TaylorT4 are taken from Tables. VII and VIII of \cite{Osh13}.  For consistency, we assume the same maximum frequencies for both waveforms, as 560 Hz and 890 Hz for the non-spinning and aligned-spin binaries, respectively.} 
  \end{table*}

%

\section{Acknowledgments} 
This study was financially supported by the ¡º2013 Post-Doc. Development Program¡» of Pusan National University. H. S. C. and C. H. L. are supported in part by the National Research Foundation Grant funded by the Korean Government (No. NRF-2011-220-C00029) and the BAERI Nuclear R \& D program (No. M20808740002) of Korea.

\appendix
\section{Dependence on the cufoff frequencies: 30 Hz versus 40 Hz and $f_{\rm LSO}$ versus $f_{\rm MECO}$} \label{AppA}
\begin{figure}[b]
\includegraphics[width=\columnwidth]{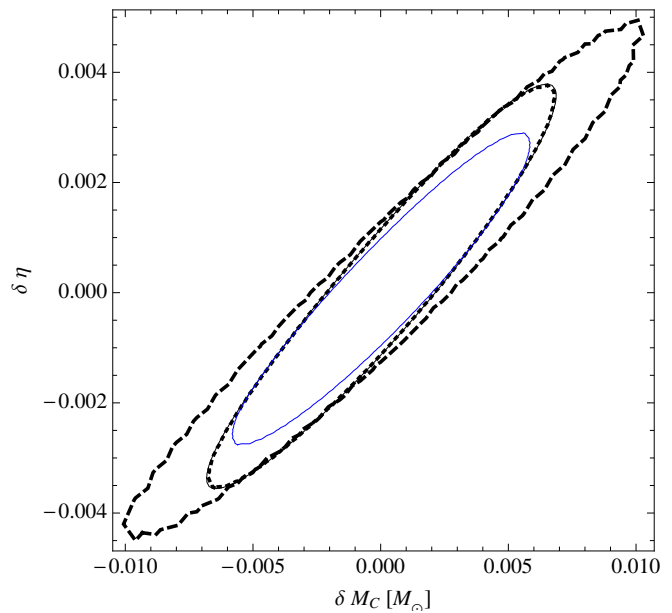}
\caption{{\bf Overlap contours ($P=0.99$) with various cutoff frequencies for the non-spinning BHNS binary.} We assume that $\{f_{\rm mim}, f_{\rm max}\}[{\rm Hz}]=\{12, f_{\rm LSO}\} ({\rm black \ solid}), \{30, f_{\rm LSO}\} ({\rm dotted}), \{40, f_{\rm LSO}\} ({\rm dashed})$, where $f_{\rm LSO}=386$ Hz, to see the dependence on the minimum frequencies, and $\{30, f_{\rm MECO}\} ({\rm blue})$, where $f_{\rm MECO}=560$ Hz,  on the maximum frequencies. There is a significant change between 40 and 30 Hz, while no change between 30 and 12 Hz (see,  Fig.~8 in \cite{Osh13} for the TaylorT4 waveform), this is because the sensitivity curve increases very rapidly below 30 Hz due to the seismic noise. The overlap also depends on the maximum frequencies. Exact numbers are summarized in Table.~\ref{tab.fcut}. \label{fig.fcut}}
\end{figure}

We show overlap contours with various cutoff frequencies using TaylorF2 waveform in Fig.~\ref{fig.fcut} and summarize the corresponding results in Table.~\ref{tab.fcut}.
To see the dependence on the minimum frequency we choose three different $f_{\rm min}$ with the same $f_{\rm max}$ as $\{f_{\rm mim}, f_{\rm max}\}[{\rm Hz}]=\{12, f_{\rm LSO}\}, \{30, f_{\rm LSO}\}, \{40, f_{\rm LSO}\}$, where $f_{\rm LSO}=386$ Hz.
The result shows a difference between 40 (dashed line) and 30 Hz (black solid line) by about $30-50$ $\%$ for the fractional errors, that are ($\Delta\mc/\mc, \Delta\eta/\eta$)=($0.08, 1.2$) for $f_{\rm min}=30$ Hz and ($0.12, 1.6$) for $f_{\rm min}=40$ Hz.
On the other hand, we do not see any difference between 30 and 12 Hz. This is because the sensitivity curve increases very rapidly below 30 Hz due to the seismic noise.  Therefore, one should choose $f_{\rm min}=30$ Hz not to lose any additional information in the frequency range lower than 40 Hz, no additional waveform is necessary lower than 30 Hz for the initial LIGO.
We consider one additional case as $\{f_{\rm mim}, f_{\rm max}\}[{\rm Hz}]=\{30, f_{\rm MECO}\}$, where $f_{\rm MECO}=560$ Hz, and find that the overlap also depends on the maximum frequency as shown in Table.~\ref{tab.fcut}.

\begin{table}[!]
\begin{tabular}{c | ccc|cc|cc  }
$f_{\max}$       &    &\multicolumn{4}{c|}{386 Hz ($f_{\rm LSO}$) } &\multicolumn{2}{c}{560 Hz ($f_{\rm MECO}$ )}    \\
 \hline
$f_{\min}$       &&\multicolumn{2}{c|}{40 Hz}  &\multicolumn{2}{|c|}{30 Hz}   &\multicolumn{2}{|c}{30 Hz}  \\
 \hline
 Parameter&&$\mc$ & $\eta$  &   $\mc$   &  $\eta$      &$\mc$ &  $\eta$ \\ 
                                                                             \hline
            $c_{ij}$                                         & $\mc$&  1.00&0.963&1.00  &0.950   &1.00&0.939   \\
                                                                   &$\eta$&-        &1.00   &-         &1.00     &-        &1.00\\
  \hline
$\sigma_i \times 10^3$		           & &       3.68  &1.70       &2.39  &1.29&   2.05  &0.997 \\
 \hline
$\Delta \sigma_i / \lambda_i \ [\%]$	 & &  0.123  &1.58 & 0.0798  &1.20      & 0.0685  &0.926    \\

    \end{tabular}
 \caption{\label{tab.fcut}{\bf Results using various cutoff frequencies for the non-spinning BHNS binary, showing the dependence on the $f_{\rm min}$ and $f_{\rm max}$.} We assume $\rho=20$. The result for $\{f_{\rm mim}, f_{\rm max}\}=\{40, f_{\rm LSO}\}$ Hz is taken from Table.~III of \cite{Cho13}, where $\rho=10$.}
 \end{table}

%
%
%


\begin{thebibliography}{9}
\bibitem{Cho13} H. -S. Cho, E. Ochsner, R. O'Shaughnessy, C. Kim, and C. -H. Lee,   {\prd} {\bf 87}, 024004 (2013).

\bibitem{Cor06} N. J. Cornish and E. K. Porter, Class. Quantum Grav. 2006, {\bf 23}, S761.
\bibitem{Van09} M. van der Sluys, I. Mandel, V. Raymond, V. Kalogera, C. R{\"o}ver, and N. Christensen, Class. Quantum Grav. 2009, {\bf 26}, 204010.
\bibitem{Cor11} N. Cornish, L. Sampson, N. Yunes, and F. Pretorius, Phys. Rev. D 84, 062003 (2011).
\bibitem{Vei12} J. Veitch, I. Mandel, B. Aylott, B. Farr, V. Raymond, C. Rodriguez, M. van der Sluys, V. Kalogera, and A. Vecchio, Phys. Rev. D 85, 104045 (2012).
\bibitem{Osh13} R. O'Shaughnessy, B. Farr,  E. Ochsner, H. -S. Cho,  C. Kim, and C. -H. Lee, (arXiv:1308.4704) (2013).


\bibitem{Poi95} E. Poisson and C. M. Will, {\prd} {\bf 52}, 848 (1995).
 \bibitem{Aru05} K. G. Arun, B. R. Iyer, B. S. Sathyaprakash, and P. A. Sundararajan, 2005, {\prd} {\bf 71}, 084008.
 \bibitem{Lan06} R. N. Lang and S. A. Hughes, 2006, {\prd} {\bf 74}, 122001.
 \bibitem{Bro07} C. Van den Broeck and A. S. Sengupta, Class. Quantum Grav. 2007, {\bf 24}, 1089.


 \bibitem{Cok08}  T. Cokelaer, Class. Quantum Grav. {\bf 25}, 184008 (2008).
\bibitem{Rod13} C. L. Rodriguez, B. Farr,  W. M. Farr, and I. Mandel,   (arXiv:1308.1397) (2013).


 \bibitem{Van08} M. van der Sluys, V. Raymond, I. Mandel, C. R{\"o}ver, V. Kalogera, R. Meyer, and A. Vecchio, Class. Quantum Grav. 2008, {\bf 25}, 184011.
\bibitem{Aas13} J. Aasi et al. (LIGO-Virgo Scientific Collaboration). {\prd} {\bf 88}, 062001 (2013).


 \bibitem{Aru09} K. G. Arun, A. Buonanno, G. Faye, and E. Ochsner, 2009, {\prd} {\bf79}, 104023. 
 \bibitem{Cut94} C. Cutler and E. \'{E}. Flanagan, {\prd}  {\bf 49}, 2658 (1994).
 

\bibitem{Sat91} B. S. Sathyaprakash and S. V. Dhurandhar,  {\prd} {\bf 44}, 3819 (1991).
\bibitem{All12} B. Allen, W. G. Anderson, P. R. Brady, D. A. Brown and J. D. E. Creighton,  {\prd} {\bf 85}, 122006 (2012).

\bibitem{Dam01} T. Damour, B. R. Iyer, and B. S. Sathyaprakash, {\prd}  {\bf 63}, 044023 (2001).
\bibitem{Aji09} P. Ajith and S. Bose, {\prd} {\bf79}, 084032 (2009).



\bibitem{Jar94} P. Jaranowski and A. Kr\'olak, 1994, {\prd} {\bf 49}, 1723.
\bibitem{Val08} M. Vallisneri, 2008, {\prd} {\bf 77}, 042001.


\bibitem{Fin92} L. S. Finn, {\prd} {\bf 46}, 5236 (1992).
\bibitem{Cut07} C. Cutler and M. Vallisneri, {\prd}  {\bf 76}, 104018 (2007).


\bibitem{lal} https://www.lsc-group.phys.uwm.edu/daswg/projects  \\ 
/lal/nightly/docs/html/


\bibitem{Buo03} A. Buonanno, Y. Chen, and M. Vallisneri, 2003, {\prd} {\bf 67}, 104025.
\bibitem{Bla02} L. Blanchet, 2002,  {\prd} {\bf 65}, 124009.


\bibitem{Lun13} A. Lundgren and R. O'Shaughnessy,   (arXiv:1304.3332) (2013).




\end{thebibliography}
\end{document}